\documentclass{article}
\usepackage{spconf}
\usepackage[utf8]{inputenc}
\usepackage{cite}
\usepackage{amsmath,amssymb,amsfonts}
\usepackage{graphicx}
\usepackage{textcomp}
\usepackage{xcolor}
\usepackage{booktabs}
\usepackage{siunitx}
\usepackage{algorithmic}
\usepackage{caption}
\usepackage{acronym}
\usepackage{url}
\usepackage[linesnumbered,ruled,vlined]{algorithm2e}
\usepackage{balance}

\DeclareMathOperator*{\argmax}{arg\,max}

\acrodef{DOA}{Direction-of-Arrival}
\acrodef{MAE}{Mean Absolute Error}
\acrodef{RIR}{Room Impulse Response}
\acrodef{LSTM}{Long short-term memory}
\acrodef{STFT}{Short-Time Fourier Transform}
\acrodef{CP}{Conformal Prediction}
\acrodef{ICP}{Inductive Conformal Prediction}
\acrodef{iRTF}{Instantaneous Relative Transfer Function}
\acrodef{DNN}{Deep Neural Network}
\acrodef{CNN}{Convolutional Neural Network}
\acrodef{ResNet}{Residual neural network}
\acrodef{ATF}{Acoustic Transfer Function}

\title{Conformal Prediction-Aware DOA Tracking}

\twoauthors
 {Maya Veisman Barness, Sharon Gannot}
	{Faculty of Engineering\\
	Bar-Ilan University, 
	Ramat-Gan, Israel\\
    \{veismam,sharon.gannot\}@biu.ac.il}
 {Bracha Laufer Goldshtein}
	{School of Electrical Engineering\\
	Tel Aviv University, Tel Aviv, Israel\\
    blaufer@tauex.tau.ac.il}

\begin{document}
\maketitle
\ninept

\begin{abstract}
We introduce an uncertainty-aware DOA-tracking framework combining deep posterior estimation, conformal-prediction calibration, and temporal smoothing. The approach enhances stability against reverberation and burst-like directional noise, yielding more reliable trajectories and greater robustness than conventional DOA-tracking methods.

\end{abstract}


\section{Introduction}
Localization of sound sources is fundamental to many modern audio applications \cite{zhu2021}, including speech enhancement and separation \cite{OferSharon2021trackSeperate} and human–robot interaction \cite{RASCON2017184}. Deep neural architectures—such as \acp{CNN}, \acp{ResNet}, and attention-based models—have demonstrated high accuracy in \ac{DOA} estimation \cite{grumiaux2022survey, liu2020nature, MACK2022101363}.
However, these models often produce uncalibrated probabilities that may be overly confident, especially in the presence of noise.

Several works incorporate tracking to exploit temporal continuity \cite{Renana2021deepRank, Biscainho2022,Bohlender2021LSTM}.
Recently, \ac{CP} has been used for uncertainty quantification in localization tasks \cite{Khurjekar2024DOAStatCov, rozenfeld2025}.



\begin{figure}[htbp]
\centering
\includegraphics[width=0.43\textwidth]{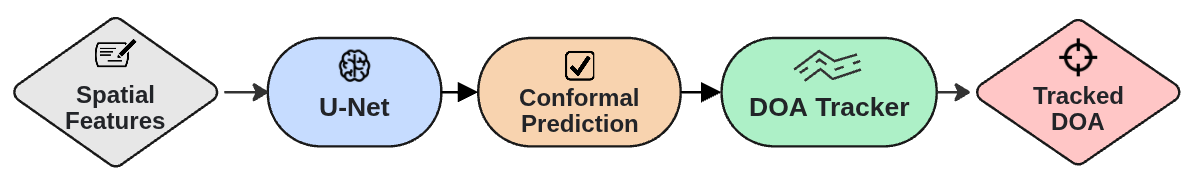}
\vspace{-8pt}
\caption{Proposed framework: iRTF extraction, U-Net posterior, conformal prediction, and DOA tracking.}
\vspace{-10pt}
\label{fig:pipeline}
\end{figure}

In this work, we integrate \ac{CP} into a \ac{DOA} tracking scheme. The proposed method includes a three-stage framework combining a \ac{DNN}-based estimator, a \ac{CP} block
\cite{angelopoulos2023, khurjekar2023uqdoa}, 
and a \ac{DOA} time-smoother. The smoother adaptively balances the influence of the neural posteriors and the temporal continuity. This balance is guided by the confidence derived from the \ac{CP} set.
The overall processing pipeline is illustrated in Fig.~\ref{fig:pipeline}.
Our main contributions are: (1) a \ac{CP} block that quantifies the per-frame reliability of a neural estimator; (2) a lightweight \ac{DOA} time-smoother weighted by the confidence measures extracted from the conformal set; and (3) experimental results showing improved accuracy and robustness under reverberation and noise bursts.
The code is publicly available.\footnote{\texttt{https://github.com/MayaVB/conformal-doa-tracking}}

\section{Methodology}
\subsection{Signal Model and iRTF Representation} 
We consider $M$ microphones capturing a single source in a noisy environment. The signals are given in \ac{STFT} domain as:
\begin{equation}
z_m(t,f) = s(t,f)\,h_m(t,f) + v_m(t,f), \quad m = 1,\ldots,M,
\end{equation}
where $t$ and $f$ denote the time-frame and frequency-bin indices, respectively. The term $h_m(t, f)$ denotes the time-varying \ac{ATF} relating the source and the $m$-th microphone, and $v_m(t,f)$ denotes an additive noise term.
The \ac{iRTF} is defined as the ratio between \acp{ATF}, and can be estimated from the microphone signals:
\begin{equation}
\mathrm{iRTF}(t,f,m) = \frac{z_m(t,f)}{z_{\mathrm{ref}}(t,f)} \approx\frac{h_m(t,f)}{h_1(t,f)}
, 
\quad m = 2,\ldots,M,
\label{iRTF_eq}
\end{equation}
Note that the \ac{iRTF} is trivial for $m=1$, which is arbitrarily selected as the reference microphone.
The complex-valued \acp{iRTF}, decomposed into their real and imaginary parts for 
$m = 2,\ldots,M$, serve as the input tensor to the \ac{DNN}-based instantaneous \ac{DOA} estimator:
\begin{equation}
\mathcal{R}(t,f,m) =
[\Re(\mathrm{iRTF}(t,f,m)),\, \Im(\mathrm{iRTF}(t,f,m))].
\label{features}
\end{equation}
\vspace{-20pt}
\subsection{Neural Network Architecture and Conformal Calibration}
\vspace{-3pt}\noindent\textbf{Neural Network Architecture.}  
The proposed DOA estimator builds upon prior work \cite{hammer2021} and is implemented as a 2D U-Net–based convolutional network that processes time–frequency representations of the multi-microphone input.\footnote{\url{https://github.com/xiaopeng-liao/Pytorch-UNet}}
We recast the \ac{DOA} estimation task as a classification problem, where the \ac{DOA} is selected from a discrete set of 
$K=32$ candidate angles: $\Theta = \{10^{\circ}, 15^{\circ}, 20^{\circ}, \ldots, 165^{\circ}\}.$
We exclude angles within $10^{\circ}$ of the endfire, where linear arrays typically suffer from reduced angular resolution compared to broadside directions.
The architecture \cite{hammer2021} follows the encoder–decoder structure with skip connections, consisting of four downsampling and four upsampling blocks with a bottleneck layer in between.
The input tensor \eqref{features} has $2(M-1)$ channels, corresponding to the real and imaginary components of the \acp{iRTF}, and dimensions $F \times T$ representing frequency and time.
The convolutional encoder outputs class-specific feature maps of size $K \times F \times T$. A fully connected layer collapses the frequency dimension to produce frame-wise softmax probabilities $\hat{p}(\theta|\mathcal{R})$, from which the \ac{DOA} estimate is obtained as $\hat{\theta}_t = \argmax_{\theta} \hat{p}(\theta \mid \mathcal{R}_t)$.

\sloppy
\vspace{2pt}
\noindent\textbf{\acf{CP}.}
\ac{CP} provides a finite-sample, distribution-free guarantee on prediction coverage. 
For a target miscoverage level $\alpha$, it ensures that
\begin{equation}
\mathbb{P}\big\{\,\theta_t \in C_t(\alpha)\,\big\} \;\ge\; 1-\alpha,
\label{eq:cp_coverage}
\end{equation}
where $\theta_t$ denotes the ground-truth \ac{DOA} class at frame $t$, 
and $C_t(\alpha)$ is the \ac{CP} set \cite{angelopoulos2023, khurjekar2023uqdoa}.

In our setup, the data is split into three subsets: a training set for neural network training, a calibration set for computing the conformal threshold, and a test set for evaluation. 
Both training and calibration sets consist of static scenes, whereas the test set consists of a moving-speaker scenario. 
Since the test data is dynamic, the exchangeability assumption in (\ref{eq:cp_coverage}) holds only approximately.
\sloppy
A nonconformity score is computed as 
$s(\theta,\mathcal{R}_t) = -\log \hat{p}_t(\theta|\mathcal{R}_t)$.
During calibration, we evaluate this score using the true label $\theta_t$, i.e., $s(\theta_t,\mathcal{R}_t)=-\log \hat{p}_t(\theta_t|\mathcal{R}_t).$
%
Let $\tau_\alpha$ be the $\lceil (1-\alpha)(n_{\mathrm{cal}}+1) \rceil$-th
order statistic of the calibration scores \cite{angelopoulos2023}, where, $n_{\mathrm{cal}}$ is the number of calibration samples. Then, the \ac{CP} set is defined as:
\begin{equation}
C_t(\alpha) = \{\theta \in \Theta : s(\theta,\mathcal{R}_t) \le \tau_\alpha \}.
\end{equation}
The set $C_t(\alpha)$ expands in uncertain or noisy conditions and shrinks when the model is confident, providing a calibrated measure of uncertainty.
\subsection{DOA Tracking with CP-Aware Weighting}
\label{subsection:DOA_Tracking_CP-Aware}
We define a row-stochastic Gaussian transition matrix\cite{bayesfilter, roweis1999unifying} $\mathbf{T} \in \mathbb{R}^{K \times K}$  over the discrete \ac{DOA} grid.
Each entry $T_{i,j}$ models the probability of moving from 
$\theta_i$ at time $t\!-\!1$ to $\theta_j$ at time $t$:
\begin{equation}
T_{i,j} =
\frac{
    \exp\!\left(-\tfrac{1}{2}\left(\tfrac{|i-j|}{\sigma_c}\right)^{2}\right)
}{
    \sum_{b=0}^{K-1} 
    \exp\!\left(-\tfrac{1}{2}\left(\tfrac{|i-b|}{\sigma_c}\right)^{2}\right)
},
\label{eq:transition_matrix}
\end{equation}
where $\sigma_c$ controls the spread of allowable transitions.
%
At each frame $t$, the smoother propagates the belief from the previous step by multiplying by the transition matrix $\tilde{\mathbf{b}}_t = \mathbf{T}\,\mathbf{b}_{t-1},$ yielding the one-step prior.  
The updated belief is obtained by blending this prior with the neural posterior $\hat{\mathbf{p}}_t$:
\begin{equation}
\mathbf{b}_t = (1 - w_t)\,\tilde{\mathbf{b}}_t + w_t\,\hat{\mathbf{p}}_t.
\end{equation}
This update yields a belief vector $\mathbf{b}_t$ whose entries represent probabilities over all \ac{DOA} classes. The final \ac{DOA} estimate is then obtained as the class with the highest probability.

The smoothness score $w_t$ is computed from three numerical descriptors extracted
from the \ac{CP} set:
\begin{equation}
w_t = \exp\!\Big[-\big(
\lambda_{\text{sp}}\,\text{var}_t
+ \lambda_{\text{sz}}\,\text{size}_t
+ \lambda_{\text{span}}\,\text{span}_t
\big)\Big].
\end{equation}
where $\text{var}_t = \mathrm{Var}(\{\theta_{i+1}-\theta_i\}),\>
1\leq i\leq|C_t|-1,\>
\text{size}_t = |C_t|$ and
$\text{span}_t = \max(C_t)-\min(C_t)$.
%
%
Small, contiguous sets increase $w_t$, whereas large or irregular ones decrease it.
%
In our experiments, we set 
$\lambda_{\text{sp}}=1$, $\lambda_{\text{sz}}=2$, and $\lambda_{\text{span}}=1$.




\begin{algorithm}[t]
\SetAlgoLined
\KwIn{
Transition matrix $\mathbf{T} \!\in\! \mathbb{R}^{K\times K}$; posteriors (from U-Net) $\{\hat{\mathbf{p}}_t\}_{t=0}^{T_{\max}}$; conformal sets $\{C_t\}_{t=0}^{T_{\max}}$; weights $\lambda_{\text{sp}}, \lambda_{\text{sz}}, \lambda_{\text{span}}$
}
\KwOut{DOA estimates $\{\tilde{\theta}_t\}_{t=0}^{T_{\max}}$}

Initialize $\mathbf{b}_0[k] \leftarrow \hat{\mathbf{p}}_0[k]$ for $k=1,\ldots,K$\\[2pt]

\For{$t=1$ \KwTo $T_{\max}$}{

  Compute $\text{var}_t$, $s_t$, and $\text{span}_t$\;

  $w_t \leftarrow \exp\!\big[-(\lambda_{\text{sp}}\text{var}_t
   + \lambda_{\text{sz}}s_t
   + \lambda_{\text{span}}\text{span}_t)\big]$\;

  $\tilde{\mathbf{b}}_t \leftarrow \mathbf{T}\,\mathbf{b}_{t-1}$\;

  $\mathbf{b}_t \leftarrow (1-w_t)\,\tilde{\mathbf{b}}_t + w_t\,\hat{\mathbf{p}}_t$\;

$\tilde{\theta}_t \leftarrow \arg\max_k\, \mathbf{b}_t[k]$\; 

}
\caption{DOA tracking with conformal-prediction-aware weighting.}
\label{alg:tracker}
\end{algorithm}

\section{Experiments}
In our experimental study, we use two datasets: a static dataset for training and calibration, and a dynamic dataset for evaluation. Both datasets share the same set of source positions along a circular trajectory of radius \SI{1.5}{m} around the array. The static dataset includes 100 simulated rooms. In each room, the source is placed sequentially at all \ac{DOA} grid positions, and a full utterance is spoken at each fixed position, yielding long frame sequences per class. We split the dataset room-wise: 80 rooms (80\%) for training and 20 rooms (20\%) for calibration.
The dynamic dataset contains 50 rooms, in which the source continuously moves along the same semi-circular path while uttering a single sentence. Therefore, each \ac{DOA} class is observed as the speaker passes through its angular region. 
This dataset is used for evaluating the proposed time smoother.

For the static dataset, \acp{RIR} were generated using gpuRIR \cite{garcia2021gpurr}, while the dynamic scenarios were produced with the AudioLabs moving-source generator.\footnote{\url{https://github.com/audiolabs/rir-generator}} adapted to the Python environment.
Clean speech was taken from the WSJ0 corpus. 
Simulations used 5-mic linear arrays in randomized shoebox rooms with $\mathrm{RT60} \in \{0.2, 0.5, 0.8\}$\,s.
During the test phase, directional burst noise was added to the target speaker signal at a local SNR of 0\,dB within each injected segment. Each burst was emitted from one of the previously visited source positions along the trajectory at a random time. Burst durations are 300–500\,ms, with 1–5 occurrences.

The U-Net classifier was trained using a cross-entropy loss and Adam optimizer with a learning rate of $10^{-4}$, weight decay $10^{-4}$, and a batch size of $30$ for $150$ epochs. In our setting, we use $\alpha = 0.1$.

Table~\ref{tab:results} reports the \ac{MAE} before and after time smoothing. In the noise-free condition, the U-Net already provides accurate and stable estimates, so smoothing yields only minor changes and may slightly degrade performance at low reverberation. Under burst noise, however, smoothing consistently reduces \ac{MAE} across all RT60 values, showing that the conformal-aware tracker effectively suppresses transient directional interference.
\begin{table}[t]
\centering
    \setlength{\belowcaptionskip}{-10pt}
\setlength{\abovecaptionskip}{3pt}
\caption{\ac{MAE} in degrees for DOA estimation across varying reverberation times (RT60), with and without burst noise. Each cell presents results as ``before/after'' smoothing. The burst noise is added at 0\,dB SNR.}
\label{tab:results}
\begin{tabular}{lccc}
\toprule
RT60 (s) & without noise & with noise \\
\midrule
0.2 & 5.81 / 6.36 & 17.24 / 14.18  \\
0.5 & 6.34 / 5.63 & 12.79 / 9.29 \\
0.8 & 7.35 / 6.67 & 10.92 / 8.38 \\
\bottomrule
\end{tabular}
\vspace{-12pt}
\end{table}
Figure~\ref{fig:spectrogram_filtering} illustrates a typical burst-noise scenario. Time smoothing suppresses spurious \ac{DOA} caused by the bursts, while preserving the underlying trajectory of the moving source, leading to smoother, stable tracking.

\begin{figure}[htbp]
\centering
\includegraphics[width=0.43\columnwidth]{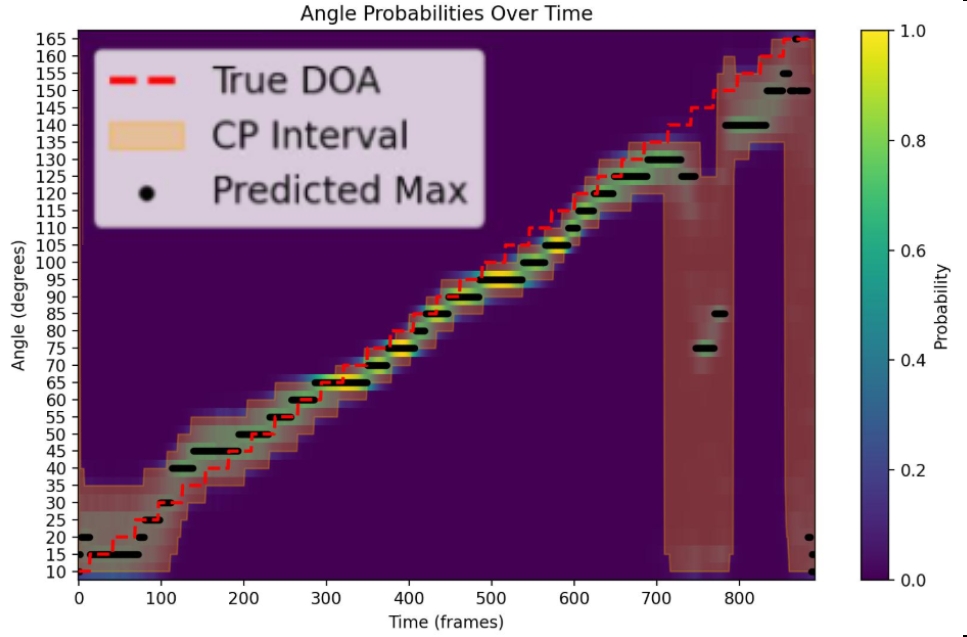}\hfil
\includegraphics[width=0.43\columnwidth]{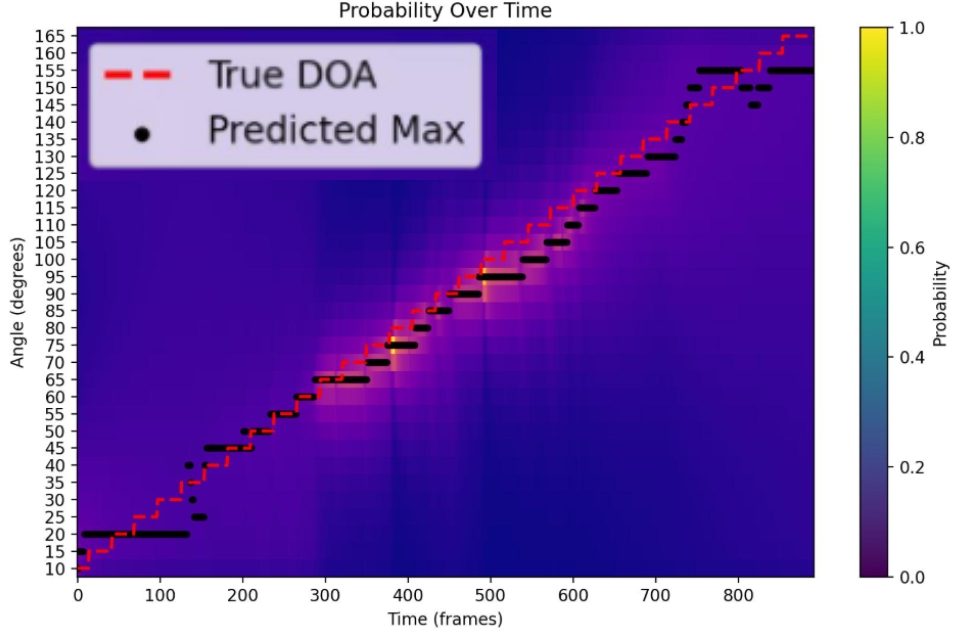}
    \setlength{\belowcaptionskip}{-12pt}
\setlength{\abovecaptionskip}{2pt}
\caption{
Frame-wise DOA probabilities before (left) and after (right) smoothing.
The speaker moves along a circular trajectory around the array at $\approx0.5$ m/s.
The conformal set expands during uncertain frames. A burst noise event after frame 800 is effectively suppressed.
}
\label{fig:spectrogram_filtering}
\end{figure}

\section{Conclusion}
We proposed a conformal-prediction-aware \ac{DOA} tracking framework that integrates a neural network, conformal calibration, and adaptive temporal smoothing. 
The proposed method improves localization robustness and temporal stability, particularly under burst-noise conditions.
\balance
\bibliographystyle{IEEEtran}
\bibliography{references}

\end{document}